\documentstyle[aps,epsfig,twocolumn]{revtex}
\begin{document}
\draft
\flushbottom
\twocolumn[
\hsize\textwidth\columnwidth\hsize\csname @twocolumnfalse\endcsname

\title{Monte Carlo determination of the phase diagram of the
double-exchange model}
\author{J. L. Alonso$^1$, J. A. Capit\'an$^2$, L. A. Fern\'andez$^2$,
F. Guinea$^3$, and V. Mart{\'\i}n-Mayor$^4$}
\address{
$^1$ Departamento de F{\'\i}sica Te\'orica, Facultad de Ciencias,
Universidad de Zaragoza, 50009 Zaragoza, Spain.\\
$^2$ Departamento de F{\'\i}sica Te\'orica, Facultad de CC. F{\'\i}sicas,
Universidad Complutense de Madrid, 28040 Madrid, Spain.\\
$^3$ 
Instituto de Ciencia de Materiales (CSIC). Cantoblanco,
28049 Madrid. Spain. \\
$^4$ Dipartimento di Fisica,
Universit\`a di Roma ``La Sapienza'', 00185 Roma and
INFN sezione di Roma, Italy.}
\date{\today}
\maketitle 
\tightenlines
\widetext
\advance\leftskip by 57pt
\advance\rightskip by 57pt

\begin{abstract}
We study the phase diagram of the double exchange model, with
antiferromagnetic interactions, in a cubic lattice both at zero and at
finite temperature. There is a rich variety of magnetic phases,
combined with regions where phase separation takes place. We identify
phases, intrinsic to the cubic lattice, which are stable for realistic
values of the interactions and dopings.  Some of these phases break
chiral symmetry, leading to unusual features.
\end{abstract}

]
\narrowtext
\tightenlines
\section{Introduction.}
The main magnetic properties of doped manganites
and other materials\cite{WK55,KS97,CVM99} are described
in terms of double exchange interactions\cite{Z51,AH55}.
An extensive body of work has been devoted to the understanding
of this model, and the related
ferromagnetic Kondo lattice,
since the pioneering work of de Gennes\cite{G60}
(for a recent review, see\cite{DHM00}).

While the model is extremely simple to formulate, as it can be 
written in terms of two dimensionless parameters 
only (see next section), the
elucidation of its phase diagram is yet to be completed.
Initial studies\cite{G60} suggested that the phase diagram
could be understood in terms of
a ferromagnetic and 
an antiferromagnetic phase with an intermediate phase
with a canted arrangement of moments. More recent work
showed that in the region of the phase diagram where
canting was expected, phase separation is more
likely\cite{N96,RHD97,AG98,Yetal98,AGG99}. 
By now, there are extensive studies
of the phase diagram of the model in one 
dimension\cite{RHD97,Yetal98,Getal00},
two dimensions\cite{Yetal98,Aetal00}, 
and infinite dimensions\cite{AGG99,F95,GGA00,HV00,CMS00}.  Even in these
relatively simple cases, the model shows a rich 
phase diagram, which is not completely elucidated.

The situation is far from understood  in the more relevant case
of three dimensions. There is a general consensus that
in the limit of infinite Hund's coupling (see next section)
the system has a ferromagnetic phase, although phase separation
near the Curie temperature at 
low dopings has also been found\cite{AGG99,TP00}.
If the Hund's coupling is not infinite, or if there are antiferromagnetic
superexchange interactions present, the situation becomes more
complicated. There is evidence that an
antiferromagnetic coupling can induce a first order
phase transition at relatively large dopings\cite{Aetal01a}.
Emphasis on the presence of first-order transitions at low temperatures
has also been done in Refs.~\cite{Metal00,Yetal00}.
In three dimensions, 
Berry's phases in the wave function of the 
electrons may arise\cite{CBV98,Aetal01b}, leading to another
source of complexity.

The present work presents a numerical study of the double
exchange model in three dimensions,
using a method discussed elsewhere\cite{Aetal01c}. We show that 
the interplay between antiferromagnetic interactions
and a three dimensional structure leads to a
rich phase diagram, where, in addition to 
ferro- and antiferromagnetic phases, other ordered
phases with cubic symmetry are possible.  The model is discussed
in the next section. Then, we present the calculation at
zero temperature and the finite temperature
results. In the last section, we discuss the main features of 
our findings, and their experimental implications.

\section{The model.}
\subsection{The Hamiltonian.}
We study the double exchange model. In its simplest form,
electrons in a cubic lattice are coupled to localized spins
defined at the same lattice sites. The spin of the electrons
is constrained to be parallel to that of the localized
spins. This restriction leads to a modulation of the electron
hopping between lattice sites, which depends on the relative
orientation of the core spins. In addition, one can define a
direct coupling between spins at different sites.
The Hamiltonian is:
\begin{equation}
\!{\cal H} = -J_{\mathrm AF}\! \sum_{ij} {\mbox{\boldmath$S$}_i}\cdot 
{\mbox{\boldmath$S$}_j}+ 
t\! \sum_{ij} \!\left(\!\langle \theta_i \phi_i | \theta_j \phi_j 
\rangle c_i^\dag c_j +  {\mathrm h. c.}\!\right). \!
\label{doublee}
\end{equation}
where $\langle \theta_i \phi_i | \theta_j \phi_j\rangle$ is the overlap
 between the spinors defined by the polar angle $\theta$, and the
 azimuthal angle $\phi$:
\begin{equation}
\langle \theta_i \phi_i | \theta_j \phi_j
\rangle = \cos\frac{\theta_i}{2} 
\cos \frac{\theta_j}{2}
+ \sin \frac{\theta_i}{2} \sin \frac{\theta_j}{2} 
e^{- i (\phi_i - \phi_j)}\,.
\end{equation}
We describe
the spins $\mbox{\boldmath$S$}$ by classical variables,
normalized to one, $|  \mbox{\boldmath$S$} | = 1$.
The system depends on two
dimensionless parameters only, the filling of the electron 
band, $x$, and the ratio $J_{\mathrm AF} / t$ (notice that our 
$J_{\mathrm AF}$ is opposite in sign to the convention 
of~\cite{DHM00})
 
The double exchange Hamiltonian, Eq.~(\ref{doublee}),
can be derived from the model which describes a lattice
of atoms with strong intra-atomic Hund's coupling between
electrons in different orbitals. The simplest 
Hamiltonian which includes this effect is that of the
ferromagnetic Kondo lattice,
\begin{equation}
{\cal H}_{\mathrm FK}=t\!\sum_{i,j,s}\! \left(c_{i,s}^\dag c_{j,s} + {\mathrm h. c.}\!\right)\!-\!
J_{\mathrm H}\!\sum_{i,s,s'} c_{i,s}^\dag  c_{i,s'}\mbox{\boldmath$\sigma$}_{ss'}\!\cdot\!
{\mbox{\boldmath$S$}_i}\,,
\label{FK}
\end{equation}
where $J_{\mathrm H} > 0$, and the $\sigma$'s are Pauli matrices.
When $J_{\mathrm H} \gg t$, the spin sub-band with an antiparallel
orientation to the core spin lies at high energy, and can be projected
out.  To first order in $1/J_{\mathrm H}$, one finds an
antiferromagnetic interaction between nearest neighbor core spins,
leading to the double exchange Hamiltonian, Eq.~(\ref{doublee}), with
$|J_{\mathrm AF}| = t^2 / ( 4 J_{\mathrm H} )$ (note that we have
normalized the core spins). This effect adds to an antiferromagnetic
superexchange interaction between the Mn ${\bf t_g}$ core
spins. Therefore, the most phenomenologically interesting range for
$|J_{\mathrm AF}|/t$ is $0.008$ --- $0.15$~\cite{Aetal01a}.

\subsection{Other interactions.}

The model discussed above is supposed to be a reasonable
starting point to the study of the magnetic properties
of doped manganites, La$_{1-x}A_x$MnO$_3$, where
$A$ stands for a divalent cation. The electrons of the
conduction band arise from the Mn ${\bf e_g}$ orbitals, while
the core spin is built up from three electrons in 
the Mn ${\bf t_g}$ orbitals. In a cubic lattice,
the ${\bf e_g}$ orbitals are doubly degenerate. 
This degeneracy is ignored in Eq.~(\ref{doublee}).
Thus, we cannot study effects associated to orbital 
ordering\cite{MKS00}. These effects are probably
important in explaining the different phases with
charge ordering observed in these materials.
The model studied here can be considered a first
approximation to the study of the magnetic properties,
in situations where there is no charge ordering and
the lattice symmetry is cubic.

The existence of double degeneracy in the 
${\bf e_g}$ orbitals implies the possibility of lattice Jahn-Teller
distortions, which have been considered the cause of the
breakdown of cubic symmetry at low dopings. In addition,
dynamic Jahn-Teller fluctuations can play a role in
some properties of doped compounds with cubic symmetry\cite{MLS95}. 
We will ignore these interactions. Previous studies\cite{Aetal01a}
suggest that the model used here suffices to understand the
main features of the magnetic transitions in the manganites,
in the doping range $0.1 \le x \le 0.4$ where most
experiments are done.

The treatment of the core spins as classical variables 
will overestimate the tendency of the system toward
long range order. Previous studies\cite{AG98} suggest that
this effect is small, as results obtained for the 
physical value $S = 3/2$ are very close to those 
found in the limit $S \rightarrow \infty$. In addition,
the only transitions at zero temperature found 
in our work are discontinuous (see next section).
Hence, quantum spin fluctuations are bounded, and cannot
change qualitatively the results.
\section{Results.}
\subsection{Phase diagram at zero temperature.}~\label{ZEROT}

The phase diagram at zero temperature can be analyzed by minimizing
the total energy, Eq. (\ref{doublee}), as function of the spin
configuration, for the different values of $x$ and the ratio $J_{\mathrm AF} /
t$.  However, previous studies in one, two and infinite
dimensions~\cite{N96,RHD97,AG98,Yetal98,AGG99,Getal00,Aetal00} have
shown that the system has a strong tendency toward
phase-separation. This enforces us to use the Maxwell construction to
correctly find the phase boundaries. The easiest way of performing the
Maxwell construction consist in minimizing the grand-canonical energy
\begin{equation}
{\cal H}^{\mathrm GC}= {\cal H} - \mu\sum_i c_i^\dag c_i\,,
\label{GRANDCANONICAL}
\end{equation}
as a function of the spin configuration for the different values of
$\mu$ and $J_{\mathrm AF}$.  In this way, we obtain the
band filling, $x$, as a function of the chemical potential $\mu/t$ and of
$J_{\mathrm AF}$. At the phase boundaries $x$ is a discontinuous (growing)
function of $\mu$: the jump at the discontinuity is the
phase-separation compositional range.

For the above mentioned minimization we will limit ourselves to the
possible spin textures observed in Monte Carlo calculations at finite
temperatures (see next subsection) and to those which seem plausible
on physical grounds.  Some of the most stable phases analyzed are
described in Table I. We have studied only the range $x\le 0.5$, due to
the particle-hole symmetry of the Hamiltonian (\ref{doublee}).

We find the ferromagnetic (FM) and antiferromagnetic (G-AFM)
phases. The A-AFM configuration is ferromagnetic within planes, and
antiferromagnetic between neighboring planes. The C-AFM configuration
is ferromagnetic along lines, and antiferromagnetic between
neighboring lines. These phases have already been discussed in the
literature. The twisted phase interpolates between the A-AFM and the
C-AFM phases. Island phases have been reported in calculations in one
and two dimensional systems\cite{Getal00,Aetal00}. These are
structures of spins aligned in one direction with the sense varying
periodically. For the values of $J_{\mathrm AF}/t$ considered here, we
only find the $(\frac{\pi}{2},\frac{\pi}{2},\pi)$ (that corresponds $2\times 2\times
1$ blocks of equal spins) and $(\frac{\pi}{3},\pi,\pi)$ ($3\times 1\times 1$
blocks), although we have tried different
combinations\cite{ISLANDS}. In these phases, the electrons are
localized, and the electronic density of states is built up of delta
functions. Helix phases are ordinary spin density waves.  It is
interesting to note that helix phases are completely irrelevant in the
thermodynamic limit, but they are remarkably stable on finite
lattices, and should be considered when analyzing numerical
simulations.

We find two new phases, intrinsic to the cubic lattice, labeled
flux and skyrmion in Table I. The spin configuration is not coplanar,
and the spins are parallel to the diagonals of the unit cube
(note that the energy is invariant under a global rotation,
so that only the relative angles are relevant). In the skyrmion
phase, the spin directions around a unit cube 
point out from the center, with a hedgehog shape.
The induced Berry phase can be thought of as generated
by an effective magnetic monopole inside each of the unit cubes which
build up the three dimensional lattice. The charge 
of the monopoles change sign in neighboring cubes,
leading to zero total charge. The flux within each cube
is isotropic, and equal to $\pi / 3$ (skyrmion configuration)
and $2\pi/3$ (flux configuration) per plaquette.
In both cases, the lattice unit contains eight sites, and
the global symmetry remains cubic.

\begin{figure}
\centerline{\epsfig{file=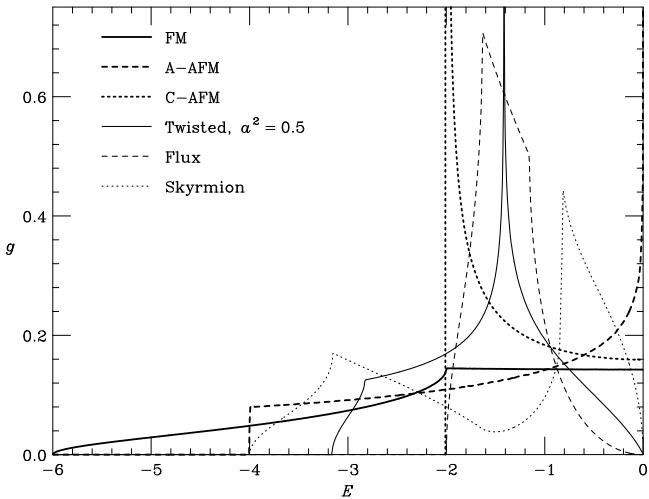,width=2.6in,angle=90}}
\caption{Density of states for several spin configurations}
\label{DOS}
\end{figure}

The electronic dispersion relation for all these phases can be
calculated analytically, and it is given in Table I.  The
corresponding density of states is shown in Fig.~\ref{DOS} . The
dispersion relation in the flux and skyrmion phases depend linearly on
$\mbox{\boldmath$k$}$ near 
$\mbox{\boldmath$k$}=(\pm\frac{\pi}{2},\pm\frac{\pi}{2},
\pm\frac{\pi}{2})$, and resembles the dispersion of Dirac massless
particles in the lattice.
The flux and twisted phases show
similarities to the phase with spins
at right angles found in two dimensions\cite{YKM98,AY00,Aetal00}.

In the twisted phase, for $a = b = 1/\sqrt{2}$, the planes
$xz$ and $yz$ have the same structure as the 2D pattern
studied in\cite{YKM98,AY00,Aetal00}. The flux phase,
on the other hand, has a dispersion relation which 
is the natural extension to 3D of the one for the
2D flux phase.
Note, however, that the flux phase in 2D is coplanar, and all
phases in the hopping elements can be rendered real by an
appropriate gauge transformation, while this does not happen
in the 3D phase studied here.

The stability of the flux and skyrmion phases arise from the
canting of the spin orientations, coupled to a shift in
the electronic density of states toward the band edges.
It is interesting to note that the electronic energy
of the flux phase is close to that of the C-AFM one, and
the energy of the skyrmion phase is also close to that of
the A-AFM phase, for the entire range of electronic
concentrations.

\begin{figure}
\centerline{\epsfig{file=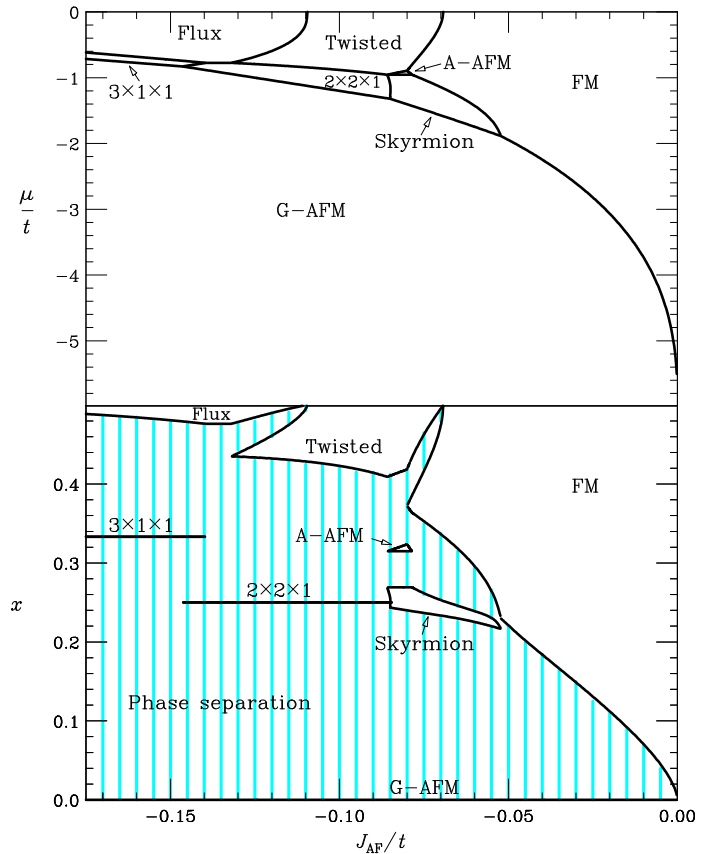,width=4.5in,angle=90}}
\caption{Phase diagram at $T=0$. Upper part: chemical potential
($\mu/t$) vs. $J_{\mathrm AF} / t$. Lower part: concentration ($x$) vs.
 $J_{\mathrm AF} / t$.
The results shown in  this plot have been obtained in a
$2400^3$ lattice. We have checked that the variations due to finite-size
effects are smaller than the line widths.
}
\label{J-MUX}
\end{figure}

The phase diagram at zero temperature is shown in Fig.~\ref{J-MUX}.
All transitions are first order. The most interesting result is the
stabilization of the skyrmion phase for a range of dopings and values
of $J_{\mathrm AF}/t$ where the model is applicable to real
materials. The twisted phase changes continuously from $a^2=0.45$
(right) to $a^2=0.77$ (left). Notice that although the A-AFM phase is
beside the twisted phase, there is a discontinuity in the $a$
parameter ($a=0$ for A-AFM).
We find island phases\cite{Aetal00} in a more restricted
range of parameters than in two dimensions, possibly due to the
competition with the flux and skyrmion phases. We have not found any
tridimensional island as
$(\frac\pi2,\frac\pi2,\frac\pi2)$. 
For more negative
values of $J_{\mathrm AF}/t$ we also find a $(\frac\pi2,\pi,\pi)$ island
phase starting at $-0.19$.

\twocolumn[
\hsize\textwidth\columnwidth\hsize\csname @twocolumnfalse\endcsname
\begin{tabular*}{\linewidth}{@{\extracolsep{\fill}}lcc}
\hline
\hline
Type&spin direction&$\epsilon(\mbox{\boldmath$k$})/t$\\
\hline & & \\
Ferromagnetic (FM) &$(0,0,1)$
&$-2\sum_{\mu=1}^3 \cos k_\mu$\\ & & \\
A-type (A-AFM)  &$(0,0,(-1)^z)$
&$-2\sum_{\mu=1}^2 \cos k_\mu$\\ & & \\
C-type  (C-AFM) &$(0,0,(-1)^{x+y})$
&$-2\cos k_3$\\ & & \\
G-type (G-AFM) &$(0,0,(-1)^{x+y+z})$
&0\\ & & \\
Twisted &$(a(-1)^{x+y},b(-1)^z,0)$
&$\pm 2\sqrt{a^2 \cos^2 k_3 + b^2 (\cos k_1 \pm \cos k_2)^2}$\\ 
&$a^2 + b^2 = 1$& \\ & & \\
Flux &$((-1)^{y+z},(-1)^{x+z},(-1)^{x+y})/\sqrt 3$
&$\pm 2\sqrt{\frac{1}{3}\sum_{\mu=1}^3 \cos^2 k_\mu}$\\ &  & \\
Skyrmion &$((-1)^x,(-1)^y,(-1)^z)/\sqrt 3$
&$\pm 2\sqrt{ \frac{2}{3}} \sqrt{\sum_{\mu=1}^3 \cos^2 k_\mu
\pm\sqrt{3 \sum_{\mu\ne\nu}\cos^2 k_\mu \cos^2 k_\nu}}$\\ & &\\
Helix &$(\cos(\omega z),\sin(\omega z),0)$
&$-2\sum_{\mu=1}^2\cos k_\mu-2\cos(\omega/2)\cos(k_3+\omega/2)$\\ & & \\
Island $\left( \frac{\pi}{2} , \pi , \pi \right)$ &
$(0,0,(-1)^{[\frac{x}{2}]+y+z})$
& $- 1 , 1 $\\ & & \\
Island $\left( \frac{\pi}{3} , \pi , \pi \right)$ &
$(0,0,(-1)^{[\frac{x}{3}]+y+z})$
& $-\sqrt{2},0,\sqrt{2}$ \\ & & \\
Island $\left( \frac{\pi}{2} , \frac{\pi}{2} , \pi \right)$ &
$(0,0,(-1)^{[\frac{x}{2}]+[\frac{y}{2}]+z})$
&$ - 2 , 0 , 0 , 2 $\\ & & \\
\hline
\hline
\end{tabular*}
\vskip 0.5cm
Table I. Spin configurations and  electronic dispersion relations of the
different phases considered in the text. The notation 
 ${[\cdot]}$ stands
for the integer part. For a lattice of linear dimension $L$, $\omega$
and $k_\mu$ can be written as $\frac{2\pi}{L}n$, with $n$ integer (in
twisted, flux and skyrmion configurations $n$ is even as the unitary
cell is a $2^3$ cube).
\vskip 0.5cm
]
\narrowtext

\subsection{Finite temperature results.} 

We have extended the previous studies to finite temperatures by using
the hybrid Monte Carlo method reported elsewhere~\cite{Aetal01c},
which allows to thermalize $16\times 16\times 16$ clusters in
reasonable computer time. We have studied the model (\ref{doublee}),
on the cubic lattice of side $L$ with periodic boundary conditions,
using the grand-canonical ensemble~(\ref{GRANDCANONICAL}). The use of
an efficient algorithm has been crucial, because around two hundred
points of the $(T,\mu,J_{\mathrm AF}/t)$ phase-diagram have been
studied.

Before presenting our results it is worth to comment on the {\em
 strong} finite-size effects of this model. Fortunately, one
can have a good analytical command on them (at least at zero
temperature), by repeating the minimization of subsection~\ref{ZEROT}
on a finite lattice (this is straightforward using Table I). In this
way one discovers that the phase boundaries change considerably with
the lattice size. Even worse, for lattice sizes $L=4$ and $8$,
the helix phase of minimal frequency $\omega=2\pi/L$ (see Table I) is
more stable than the ferromagnetic phase in important regions of the phase
diagram. Also the C-AFM phase is stable in the $L=8$, 
although it is unstable in the $L\to\infty$ limit. On
the other hand, the zero temperature phase diagrams of the $L=6$ and
$L=10$ lattices are rather close to the infinite volume limit. The
only caveat of this lattice sizes is that they cannot accommodate
island-phase whose unit cell is not commensurate with them (in
particular, the $(\frac{\pi}{2},\frac{\pi}{2},\pi)$ island phase, 
that has a $4\times
4\times 2$ unit cell). We have therefore chosen to perform most of
the numerical simulations on the $6^3$ cluster and, only when
necessary for elucidation, we have employed as a larger cluster the
$10^3$. Exception to this rule has been the simulation of the $8^3$ in
the range $-0.29\le J_{\mathrm AF}/t\le -0.18$ range, to check for the
presence of the $2\times 2\times 1$ island phase.

\begin{figure}
\centerline{\epsfig{file=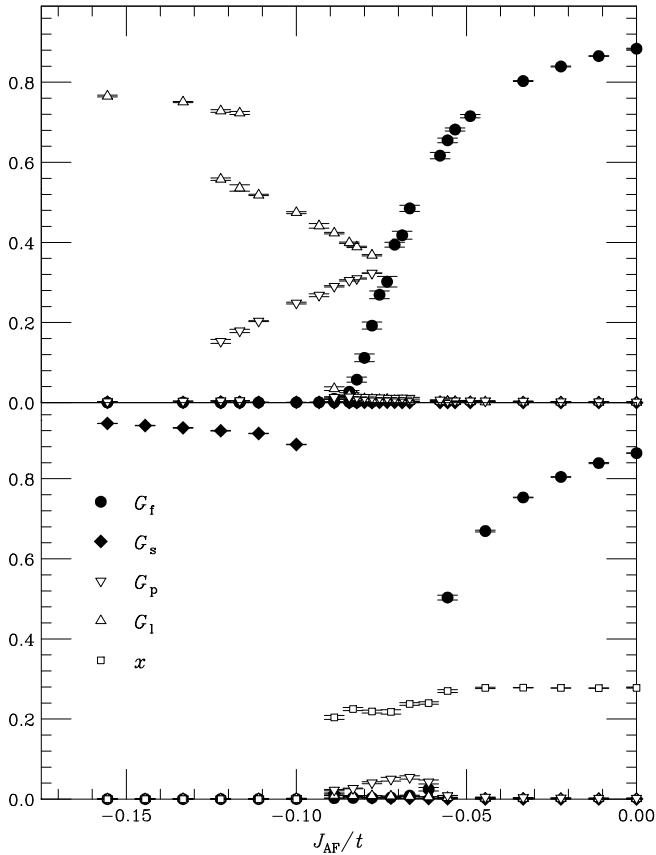,width=4.5in,angle=90}}
\caption{Monte Carlo measures of several quantities at $T=t/45$ for
$\mu=0$ (upper side) and $\mu=-1.5$, on the $L=6$ lattice.}
\label{MCMAG}
\end{figure}

In the simulation, we have used the {\em perfect action}
formulation~\cite{Aetal01c}, with $\lambda=0.25$. 
The molecular dynamics parameters have been chosen to
ensure an acceptance rate greater than an 80\%. The selection depends 
mainly on the temporal lattice size (inverse temperature).
Typical values used are  $N_\tau=20$, $\Delta\tau=0.02$ for $T/t=1/20$, and
$N_\tau=50$, $\Delta\tau=0.01$ for  $T/t=1/45$.
We have systematically controlled the correct thermalization of
our data, by comparing the results of hot and cold starts. In some
cases, in particular at $T=t/100$, we have also compared the results
of ordered starts with a slow annealing from high temperatures. 
The total run-length (up to 2000 trajectories in some
points) has always been at least larger
than three times the thermalization time. The typical
size of our statistical errors can be found in figure~\ref{MCMAG}. The
total CPU time devoted to this project has been the equivalent of 10
years of a Pentium III at 800 MHz. 

As we have said in the introduction, our scope has been the
determination of the magnetic phase diagram of the model.  Since one
cannot break a symmetry on a finite lattice, it is necessary to use
pseudo order-parameters for obtaining the phase
diagram~\cite{FOOTNOTE2}.  We have defined our pseudo order-parameters
in terms of the spin structure factor ($V=L^3$ is the total number of
spins),
\begin{eqnarray}
G(\mbox{\boldmath$k$})&=&\left\langle 
|\hat{\mbox{\boldmath$S$}}(\mbox{\boldmath$k$})|^2
\right\rangle\,,\\ 
\hat{\mbox{\boldmath$S$}}(\mbox{\boldmath$k$})
&=&\frac{1}{\sqrt{V}}\sum_j {\mbox{\boldmath$S$}_j} 
{\mathrm e}^{{\mathrm i} 
\mbox{\scriptsize \boldmath$k$}\cdot\mbox{\scriptsize \boldmath$r$}_j}\,.
\end{eqnarray}
In the above equation, $\langle\cdot\rangle$ stands for the
thermal average. In a paramagnetic phase, $G(\mbox{\boldmath$k$})$
will be of order $1/V$ for all $\mbox{\boldmath$k$}$. On the other
hand, on an ordered phase it will be of order one for a very small set
of $\mbox{\boldmath$k$}$'s (see below), and of order $1/V$ for the
rest of wave numbers. If one now calculates the structure factor for
the configurations in Table I, it follows that the appropriate pseudo 
order-parameters are:
\begin{eqnarray}
G_{\mathrm f}&=&G(0,0,0),\label{GF}\\
G_{\mathrm s}&=&G(\pi,\pi,\pi),\label{GS}\\
G_{\mathrm l}&=&G(\pi,\pi,0)+G(\pi,0,\pi)+G(0,\pi,\pi),\label{GP}\\
G_{\mathrm p}&=&G(\pi,0,0)+G(0,\pi,0)+G(0,0,\pi),\label{GL}\\
G_{\mathrm h}&=&\textstyle G(\frac{2\pi}{L},0,0)+G(0,\frac{2\pi}{L},0)
        +G(0,0,\frac{2\pi}{L}).\label{GH}
\end{eqnarray}
Let us explain them in details.  In a ferromagnetic phase, only
$G_{\mathrm f}$ is of order 1, while for a G-AFM phase, the
appropriate pseudo order-parameter would be $G_{\mathrm s}$. For an
A-AFM phase, the selected order parameter will be $G_{\mathrm p}$ (we
sum for $\mbox{\boldmath$k$}=(\pi,0,0),(0,\pi,0)$ and
$(0,0,\pi)$ because we do not know a priori the direction of the AFM
lattice planes).  A skyrmion phase will also have $G_{\mathrm p}$ as
pseudo order-parameter, and so the question arise of how to
differentiate a skyrmion from an A-AFM phase. The easiest way to do
that is to control that for the skyrmion phase the three quantities
$G(\pi,0,0)$, $G(0,\pi,0)$ and $G(0,0,\pi)$ are {\em
simultaneously} large (we have also numerically checked the
orthogonality of the Fourier transformed spin field at the three
$\mbox{\boldmath$k$}$ values). Also the C-AFM phase and the flux
phase have a common pseudo order-parameter, namely $G_{\mathrm l}$,
but one can differentiate both phases in exactly the same way. A twisted
phase will have as pseudo order-parameter {\em both} $G_{\mathrm p}$
and $G_{\mathrm l}$.  One can distinguish the twisted phase from a
tunneling phenomena between an A-AFM and a C-AFM phases, checking
again that both pseudo order-parameters are simultaneously large (we
have also checked the orthogonality of the Fourier transformed
spin-field at the two $\mbox{\boldmath$k$}$ for which the structure
factor is not negligible). Finally the helix phase is signaled by the
$G_{\mathrm h}$ order parameter, while the structure factor for the
$(k_1,k_2,k_3)$ island phase is heavily peaked at
$\mbox{\boldmath$k$}=(k_1, k_2, k_3)$.

The results for the $G_\alpha$ pseudo order-parameters and for the
equilibrium fermionic density at $T/t=1/45$ are shown in Fig.~\ref{MCMAG}, for
$\mu = 0$ (half filling, $x = 0.5$) and $\mu = -1.5 t$ which includes
the interesting region of $x \approx 1/4$ --- $1/3$. The numerical results
shown in this plot allow us to identify the FM, G-AFM, twisted and
flux phases, and the onset of the skyrmionic phase.

For $\mu=0$, the flux-twisted transition is clearly discontinuous. In
addition, there is a clear metastable behavior (the duplicated points
correspond respectively to two metastable phases). At the right end of
the twisted phase, $G_{\mathrm p}$ and $G_{\mathrm l}$ vanish abruptly
and there are clear signs of metastability with a phase where
$G_{\mathrm f}$ is small but non-vanishing. To check if it is a finite
size effect, we have performed some simulations in this region with a
larger lattice ($10^3$), obtaining that $G_{\mathrm f}$ is much smaller
(and $G_{\mathrm p}$ and $G_{\mathrm l}$ vanish). Thus we conclude
that there is a first order transition at this temperature between the
twisted and the paramagnetic phases. On the contrary, the
ferromagnetic-paramagnetic transition is smoother. We recall that in
Ref. \cite{Aetal01c} it is shown that at $J_{\mathrm AF}=0$,
$T/t=0.14$ the transition is second order.  Although we find a
strengthening of this transition when $|J_{\mathrm AF}|$ grows, to
confirm that it actually becomes of the first-order would require a
careful finite-size scaling analysis that is beyond the scope of this
work.

\begin{figure}
\centerline{\epsfig{file=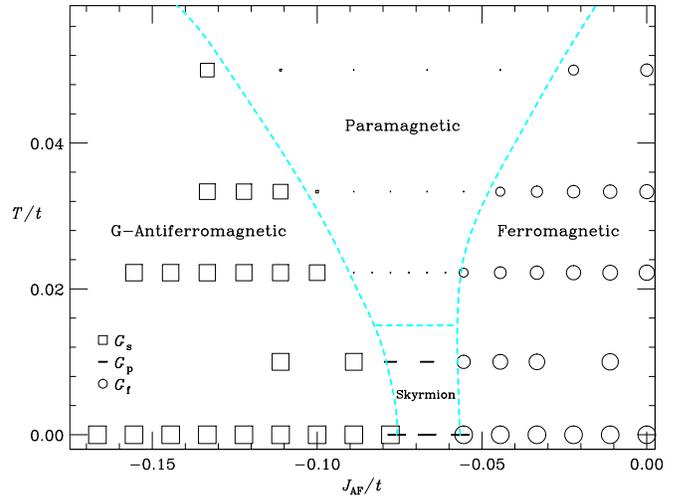,width=2.6in,angle=90}}
\caption{Phase diagram obtained with Monte Carlo measures for
$\mu=-1.5$.}
\label{MU1p5}
\end{figure}

For $\mu=-1.5$ (see Fig.~\ref{MCMAG}) we find a very different
landscape. At the left part we find an almost saturated G-AFM region
(notice that $x\approx 0$ and the fermions essentially play no
role). In the center, we find a region with $x\approx0.25$ where
$G_{\mathrm p}$ is about 0.05 and the others are much smaller. We
interpret this result as the onset of the skyrmionic region (that
appears clearly at lower temperatures, see Fig.~\ref{MU1p5}). The
start of the ferromagnetic region is much abrupter than at $\mu=0$,
but as said above, to conclude that it is first order, a study in
different lattice sizes is mandatory.

\begin{figure}
\centerline{\epsfig{file=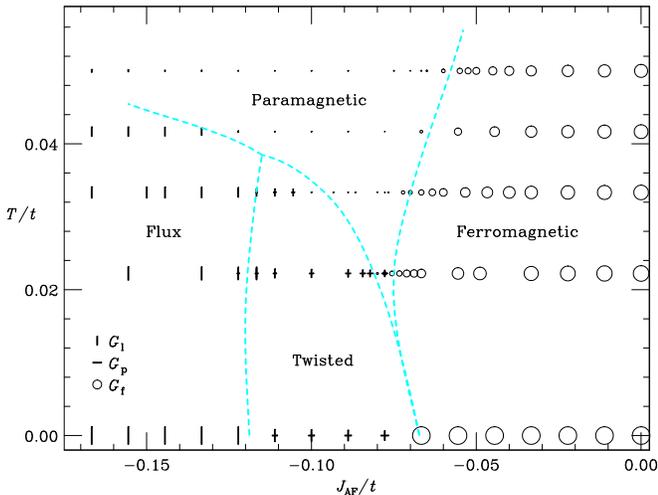,width=2.6in,angle=90}}
\caption{Phase diagram obtained with Monte Carlo measures for
$\mu=0$. The length of the segments, or the diameter of the circles are
proportional to the associated $G_\alpha$. For clarity at each point we plot
only the maximum $G_\alpha$ except in the cases where
$G_{\mathrm p}/G_{\mathrm l}\in [0.25,1]$, that correspond to the
twisted phase.}
\label{MU0}
\end{figure}

To explore the phase diagram, we have simulated at several values of
$T$.  The phase diagram at half filling is plotted in Fig.~\ref{MU0}.
The paramagnetic region separates the ferromagnetic
from the twisted phase for $T/t\ge 1/45$ but it is difficult to know 
the point where it finishes (maybe $T=0$). As the ferromagnetic
magnetization goes to zero rather smoothly, to obtain a more precise
value of the transition point we have used data from larger lattices
(mainly $L=10$). The dashed lines have been draw approximately at the
phase transitions.

Results for the different $G_\alpha$ order parameter for $\mu = -1.5 t$ are
shown in Fig.~\ref{MU1p5}. Skyrmion correlations are suppressed at
relatively low temperatures, but for $T=t/100$ we find a clear
skyrmionic structure ($G_{\mathrm p}=0.7$ --- $0.8$, with the appropriate
orthogonality of the Fourier components).
\begin{figure}
\centerline{\epsfig{file=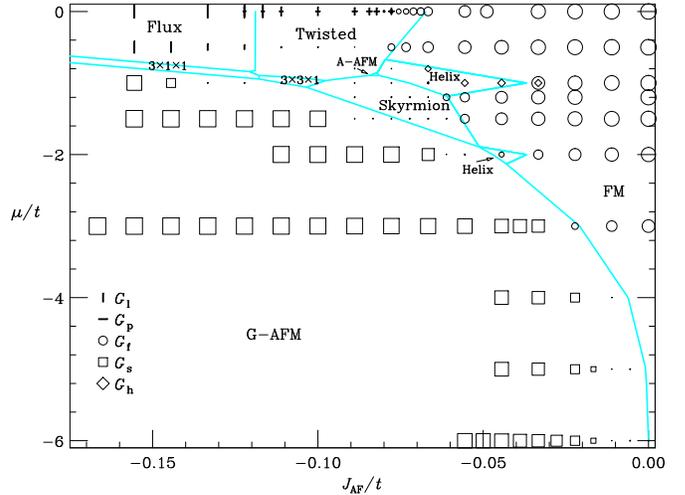,width=2.6in,angle=90}}
\caption{Phase diagram at $T=t/45$. The solid lines correspond to
$T=0$ for a $L=6$ lattice.}
\label{T90}
\end{figure}

We have combined all our results at $T=t/45$ in the phase diagram
shown in Fig.~\ref{T90}. The results are consistent with
the zero temperature calculation for the $6^3$ lattice, whose
transitions are plotted as continuous lines. 
The FM, G-AFM and Flux phases (and the twisted one near half filing) 
are less affected by finite temperature
effects than the other phases, that become paramagnetic for
temperatures as small as $T=t/45$. Notice also that in the $6^3$ lattice 
the island phases are so narrow that they are hardly
observed in a simulation. However, we have checked that in a $8^3$
lattice at $\mu=0$, $T=t/45$ the system is in the $2\times 1\times 1$ 
phase for $J_{\mathrm AF}/t\in (-0.29,-0.18)$. We have also simulated
a $8^3$ lattice 
in several values of $(\mu,J_{\mathrm AF})$ that at $T=0$ correspond to
the  $2\times 2\times 1$ region, observing that even at $T=t/45$ the
system is disordered.

\subsection{The skyrmion phase.}
As mentioned above, we find a novel phase, the skyrmion
phase, for a realistic range of dopings and antiferromagnetic
couplings. We now discuss some physical features of this phase.
Similar properties can be expected 
in the flux phase.

i) The symmetry of the magnetic phase is cubic.
This phase is compatible with experiments done in the
range $x \approx 1/4$ --- $1/3$ which show phases with cubic structure
but no macroscopic magnetization\cite{Metal97}.

ii) The magnetic arrangement is, at first sight, similar to
that expected in a canted phase. Measurements of
the microscopic magnetic structure will probably be indistinguishable
from those expected for a canted phase.

iii) The magnon spectrum will resemble that of other systems
with non collinear spins\cite{JG89,M95}. The dispersion relation
will be linear at low energies.

iv) The phase is metallic. However, the chemical potential,
at the appropriate dopings,
coincides roughly with the position when the density of states
has a pronounced minimum (see Fig.~\ref{DOS} ). Hence, it cannot be 
ruled out that residual disorder will induce  
insulating behavior. If this is the case, it will be 
behave in a similar way to the canted insulating phases
reported in the literature (see, for
instance \cite{Tetal96}).

v) The unusual spin configuration leads
to an effective coupling between an external
magnetic field and a charge density wave.
An applied field tilts the spins.  The sign of the change in the angle
between neighboring spins depends on the relative orientation
of the spins and the magnetic field. A reduction in the angle
leads to an increase of the hopping, and the opposite happens
if the angle is increased. Hence, electron charge will
be displaced  
from the weak bonds to the strong ones, and a charge
density wave will arise. This feature can clearly
identify the skyrmion phase.
 
\section{Conclusions.}

We have analyzed the phase diagram of the double exchange model in a
cubic lattice, both at zero and at finite temperature.  Our results at
finite temperature have been obtained using an efficient Hybrid Monte
Carlo algorithm~\cite{Aetal01c}, which has been crucial for the
exploration of a {\em three dimensional} phase diagram.  We have found
that the competition between the ferromagnetic interaction mediated by
the conduction electrons and antiferromagnetic couplings which arise
from superexchange effects or from the underlying intra-atomic Hund's
coupling lead to the existence of a variety of phases.

We have expressed our results in terms of the electron hopping, $t = W
/ 12$, where $W\sim 2$\,eV is the width of the conduction band, and
the effective antiferromagnetic coupling, $J_{\mathrm AF}$, arising
from superexchange interactions between the Mn ${\bf t_g}$ orbitals
($|J_{\mathrm AF}|/t \sim 0.005$ --- $0.012$~\cite{PGJ98}) and from the
finite value of the intra-atomic Hund coupling, $J_{\mathrm H}$. It
follows that the most phenomenologically interesting range for 
$|J_{\mathrm AF}|/t$ is $0.008$---$0.15$~\cite{Aetal01a}. Using
the previous values, we find that the highest transition temperature
($x=0.5,J_{\mathrm AF}=0$) is close to $350$ K, and therefore most
of the phase transitions reported in this work occur in the $0$---$300$K
range, in agreement with experiments.

It is somewhat surprising that
conventional spin density wave structures are not
stable. The three dimensional phase diagram shares
some features with solutions of
the double exchange model in one,
two and infinite dimensions, like phase separation.
A number of phases, however, have no counterparts
in other dimensions.

We have identified, among others, a phase where the spins
are locally arranged in a hedgehog manner, labelled
skyrmion phase. 
The existence of this structure
is probably intrinsic to the double exchange mechanism,
and cannot be realized in spin systems with
short range couplings. 
The skyrmion phase is stable for dopings
$x \approx 0.2$ --- $0.3$ and antiferromagnetic couplings
$|J_{\mathrm AF}| / t \approx 0.05$ --- $0.07$. These values
are realistic for doped manganites in the range
where Colossal Magnetoresistance effects are found.
While it is probably difficult to characterize this
phase experimentally, it will be interesting to
verify its existence.

\section{Acknowledgements.}

We want to warmly thank Victor Laliena for a pleasant and fruitful
collaboration. It is also a pleasure to thank B. Alascio, A. Aligia,
E. Dagotto and B. Normand, for discussions on this problem.

We acknowledge financial support from grants PB96-0875, AEN97-1680,
AEN97-1693, AEN99-0990 (MEC, Spain) and (07N/0045/98)
(C. Madrid). V.M.-M. is supported by the Ministerio de Educaci\'on y
Cultura.  The simulations have been carried out in RTNN computers at
Zaragoza and Madrid.


\begin{references}
\bibitem{WK55}
E. D. Wollan and W. C. Koehler,
Phys. Rev. {\bf 100}, 545 (1955).
\bibitem{KS97}
D. I. Khomskii and G. Sawatzky, Solid State Commun.
{\bf 102}, 87 (1997).
\bibitem{CVM99}
J. M. D. Coey, M. Viret and S. von Molnar, Adv. in Phys.
{\bf 48}, 167 (1999).
\bibitem{Z51}
C. Zener, Phys. Rev. {\bf 82}, 403 (1951). 
\bibitem{AH55}
P. W. Anderson and H. Hasegawa,  Phys. Rev. {\bf 100}, 675
(1955).
\bibitem{G60}
P.-G. de Gennes,  Phys. Rev. {\bf 118}, 141 (1960).
\bibitem{DHM00}
E. Dagotto, T. Hotta and A. Moreo, preprint
(cond-mat/0012117).
\bibitem{N96}
E. L. Nagaev,  Physica B {\bf 230-232}, 816 (1997).
\bibitem{RHD97}
J. Riera, K. Hallberg and E. Dagotto, Phys. Rev. Lett.
{\bf 79}, 713 (1997).
\bibitem{AG98}
D. P. Arovas and F. Guinea, Phys. Rev. B {\bf 58}, 9150 (1998)
\bibitem{Yetal98}
S. Yunoki, J Hu, A. L. Malvezzi, A. Moreo,
N. Furukawa and E. Dagotto,  Phys. Rev. Lett. {\bf 80}, 845 (1998).
\bibitem{AGG99}
D. Arovas, G. G\'omez-Santos and F. Guinea,
Phys. Rev. B {\bf59}, 13569 (1999).
\bibitem{Getal00}
D. J. Garc{\'\i}a, K. Hallberg, C. D. Batista, M. Avignon and
B. Alascio,  Phys. Rev. Lett. {\bf 85}, 3720 (2000).
\bibitem{Aetal00}
H. Aliaga, B. Normand, K. Hallberg, M. Avignon and B. Alascio,
preprint (cond-mat/0011342).
\bibitem{ISLANDS}
We use the notation 
$(\frac{n_1\pi}{m_1},\frac{n_2\pi}{m_2},\frac{n_3\pi}{m_3})$ that
corresponds to the frequency associated to the maximum of the Fourier 
transform. Thus,
$(\frac\pi2,\frac\pi2,\pi)$ and $(\frac\pi3,\pi,\pi)$ represent islands of
sizes $2\times 2\times 1$ and $3\times 1\times 1$ respectively. 
We have studied all frequencies with the restrictions 
$m_i\le4$, $m_1 m_2 m_3\le 9$.
In addition, we look at the phase $(\frac\pi2,\pi,\pi)+(\pi,\pi,\pi)$
(see Ref. \cite{Aetal00} for the two dimensional case)
that we find stable for  $J_{\mathrm AF}/t$ values smaller than $-0.3$
far away from the region of physical interest.
\bibitem{F95}
N. Furukawa, J. Phys. Soc. Jap. {\bf 64}, 2734 (1995);
 {\em ibid} {\bf 64}, 2754 (1995).
\bibitem{GGA00}
F. Guinea, G. G\'omez-Santos and
D. P. Arovas, Phys. Rev.B {\bf62}, 391 (2000).
\bibitem{HV00}
K. Held and D. Vollhart,  Phys. Rev. Lett. {\bf 84}, 5168 (2000).
\bibitem{CMS00}
A. Chattopadhyay, A. J. Millis and S. Das Sarma, Phys. Rev. B
{\bf 61}, 10738 (2000).
\bibitem{TP00}
N.-H. Tong and F.-C. Pu, Phys. Rev. B {\bf 62}, 9425 (2000).
\bibitem{Aetal01a}
J. L. Alonso, L. A. Fern\'andez, F. Guinea, V. Laliena and V.
Mart{\'\i}n-Mayor, Phys. Rev. B {\bf 63}, 64416 (2001).
\bibitem{Metal00} 
A. Moreo, M. Mayr, A. Feiguin, S. Yunoki, and E. Dagotto, 
Phys. Rev. Lett. {\bf 84}, 5568(2000)  
\bibitem{Yetal00} 
S. Yunoki, T. Hotta, and E. Dagotto, Phys. Rev. Lett. {\bf 84}, 3714 (2000).
\bibitem{CBV98}
M. J. Calder\'on, L. Brey and J. A. Verg\'es, Phys. Rev. B
{\bf 58}, 3286 (1998).
\bibitem{Aetal01b}
J. L. Alonso, L. A. Fern\'andez, F. Guinea, V. Laliena and V.
Mart{\'\i}n-Mayor, Phys. Rev. B {\bf 63}, 54411 (2001).
\bibitem{Aetal01c}
J. L. Alonso, L. A. Fern\'andez, F. Guinea, V. Laliena and V.
Mart{\'\i}n-Mayor, cond-mat/0007450, Nucl. Phys. {\bf B}, in press.
\bibitem{MKS00}
T. Mizokawa, D. I. Khomskii and G. A. Sawatzky,
Phys. Rev.B {\bf 61}, R3776 (2000); {\em ibid} {\bf 63}, 024403 (2001).
\bibitem{MLS95}
A. J. Millis, P. B. Littlewood, and B. I. Shraiman,
Phys. Rev. Lett. {\bf 74}, 5144 (1995).
\bibitem{YKM98}
M. Yamanaka, W. Koshibae and S. Maekawa, Phys. Rev. Lett.
{\bf 81}, 5604 (1998).
\bibitem{AY00}
D. F. Agterberg and S. Yunoki, Phys. Rev. B {\bf62}, 13816 (2000).
\bibitem{FOOTNOTE} 
In small finite lattices,
the stable island phases, with $4^3$ or  $6^3$ unit cell
appear only when compatible with the lattice size.
\bibitem{FOOTNOTE2} 
In a finite lattice, the Monte Carlo average of a {\em real} order
parameter like, for instance, the magnetization is zero also in a ferromagnetic
phase. This is due to global rotations of the spins (global rotations
cost no energy), that averages out the magnetization even if the {\em
instantaneous} magnetization is large.
\bibitem{Metal97}
J. M. De Teresa, C. Ritter, M. R. Ibarra,
P. A. Algarabel, J. L. Garc{\'\i}a-Mu\~noz, J. Blasco,
J. Garc{\'\i}a and C. Marquina, Phys. Rev. B {\bf 56}, 3317 (1997).
\bibitem{JG89}
Th. Jolicoeur and J. C. Le Guillou, Phys. Rev. B
{\bf 40}, 2727 (1989).
\bibitem{M95}
A. Mattsson, Phys. Rev. B {\bf 51}, 11574 (1995).
\bibitem{Tetal96}
Y. Tomioka, A. Asamitsu, H. Kuwahara, Y. Moritomo and Y. Tokura,
Phys. Rev.B {\bf 53}, R1689 (1996).
\bibitem{PGJ98}
I. Panas, R. Glatt and T. Johnson, J. Phys. Chem. Solids
{\bf 59}, 2230 (1998).


\end{references}
\end{document}